\documentclass[aps,prl,amsmath,twocolumn,superscriptaddress,showpacs,floatfix]{revtex4}
\usepackage{graphicx}

\begin{document}

\title{Magnetic Correlations and the Anisotropic Kondo Effect in 
Ce$_{1-x}$La$_x$Al$_3$}

\author{E. A. Goremychkin}
\affiliation{Argonne National Laboratory, Argonne, IL 60439-4845, USA}
\author{R. Osborn}
\affiliation{Argonne National Laboratory, Argonne, IL 60439-4845, USA}
\author{B. D. Rainford}
\affiliation{Department of Physics, University of Southampton, Southampton, 
SO17 1BJ, United Kingdom}
\author{T. A. Costi}
\affiliation{Universit\"{a}t Karlsruhe, Institut f\"{u}r Theorie der
Kondensierten Materie, 76128 Karlsruhe, Germany}
\affiliation{Institut Laue Langevin, F-38042 Grenoble C{\'e}dex, 
France}
\author{A. P. Murani}
\affiliation{Institut Laue Langevin, F-38042 Grenoble C{\'e}dex, 
France}
\author{C. A. Scott}
\affiliation{Department of Physics, University of Southampton, Southampton, 
SO17 1BJ, United Kingdom}
\author{P. J. C. King}
\affiliation{ISIS Pulsed Neutron and Muon Facility, Rutherford Appleton Laboratory, Chilton, Didcot, OX11 0QX, United Kingdom}

\date{\today}

\begin{abstract}
By combining the results of muon spin rotation and inelastic neutron
scattering in the heavy fermion compounds Ce$_{1-x}$La$_x$Al$_3$ ($0.0
\leq $x$ \leq 0.2$), we show that static magnetic correlations are
suppressed above a characteristic temperature, T$^*$, by electronic
dissipation rather than by thermal disorder.    Below T$^*$, an energy
gap opens in the single-ion magnetic response in agreement with the
predictions of the Anisotropic Kondo Model.  Scaling arguments suggest
that similar behavior may underlie the ``hidden order" in URu$_2$Si$_2$.
\end{abstract}

\pacs{71.27.+a, 75.40.Cx, 76.75.+i, 78.70.Nx}

\maketitle

Although CeAl$_3$ was the first material to be designated a heavy
fermion compound~\cite{Andres:1975}, the nature of the low temperature
ground state has never been
resolved~\cite{Bredl:1984,Barth:1989,Wong:1992,Gavilano:1995}.  An
anomaly in the electronic specific heat coefficient, $\gamma$, at
T$^*\approx 0.5$~K, was shown by muon spin rotation ($\mu$SR)
experiments to coincide with the development of static magnetic
correlations \cite{Barth:1989}, but no evidence for magnetic order was
found by neutron diffraction.  It is possible that the ordered moment is
too small to be observable by neutron scattering from polycrystalline
samples.  If so, CeAl$_3$ would be reminiscent of other heavy fermion
compounds exhibiting ``hidden order'', such as
URu$_2$Si$_2$~\cite{Shah:2000}, in which the magnetism is too weak to
explain the entropy associated with the observed transition.

The thermodynamic anomaly that is observed in pure CeAl$_3$ grows in
both temperature and magnitude with lanthanum doping. T$^*$ increases to
2.2~K in Ce$_{1-x}$La$_x$Al$_3$ when $x=0.2$, and there is a substantial
peak in the specific heat with an entropy approaching
$R\ln2$~\cite{Andraka:1995}. Nevertheless, in an earlier study, we were
still unable to observe magnetic Bragg peaks in high intensity neutron
diffraction~\cite{Goremychkin:2000}.  Instead, we proposed that the
specific heat peak results from a crossover in the single-ion dynamics
that is explained by the Anisotropic Kondo Model
(AKM)~\cite{Costi:1998}.  In highly anisotropic systems, such as
Ce$_{1-x}$La$_x$Al$_3$, the AKM predicts that, at high temperature,  the
spin dynamics will be purely relaxational but that, at low temperature,
an energy gap, representing a tunneling transition between two
anisotropically hybridized states, will open up in the magnetic
response.  It is the development of this spin gap that produces the
observed specific heat anomaly at T$^*$.  Various scaling relations
predicted by the AKM were in good agreement with our neutron data.

Although we were able to conclude that the spin dynamics were consistent
with the AKM~\cite{Goremychkin:2000}, we had no explanation for the
presence of static magnetic correlations below T$^*$ seen in $\mu$SR.   
Furthermore, Pietri \textit{et al}~\cite{Pietri:2001} concluded
that the field-dependence of the specific heat peaks, though unusual,
was inconsistent with the AKM and probably signified a magnetic
transition~\footnote{We believe that the discrepancies reported in
Ref.~\cite{Pietri:2001} are due to the neglect of crystal field effects
in their calculations rather than a failure of the AKM. We will address
this in more detail in a future publication.}.  It is important
to clarify the link between the development of static magnetic
correlations and the appearance of an energy gap in the spin dynamics,
in order to establish the origin of the thermodynamic anomalies and to
assess the validity of a single-ion explanation such as the AKM.

In this paper, we present new $\mu$SR and inelastic neutron scattering
results on the Ce$_{1-x}$La$_x$Al$_3$ series, with $x$ = 0.0, 0.05, 0.1,
and 0.2, that demonstrate unambiguously that magnetic order is not
responsible for the specific heat maxima.  For low values of $x$, we
observe with muons a well defined magnetic order parameter, but T$^*$
does not mark the temperature at which the magnetic moments become
thermally disordered, as in a conventional magnetic phase transition.
Rather, T$^*$ is the temperature at which the magnetic order is
suppressed by electronic dissipation.  Moreover, there is no correlation
between the temperature dependence of the spin gap and the magnetic
order parameter, as there must be at a classical magnetic phase
transition.  We discuss similarities with URu$_2$Si$_2$, another highly
anisotropic compound, in which the magnetic order appears to be a
parasitic phenomenon~\cite{Matsuda:2001}, but with a N{\'e}el
temperature T$_N$ slightly less than the ``hidden order" transition,
\textit{i.e.}, T$_N \leq$ T$^*$ in the case of URu$_2$Si$_2$, whereas
T$_N >$ T$^*$ in Ce$_{1-x}$La$_x$Al$_3$ for $x<0.2$.

The samples of Ce$_{1-x}$La$_x$Al$_3$ with $x$ = 0.0, 0.05, 0.1, and 0.2
were prepared by arc melting stoichiometric quantities of the
constituent elements, followed by annealing at 850$^\circ$C for four
weeks.  Neutron diffraction confirmed that all sample were single phase.
 The $\mu$SR measurements were performed at the ISIS pulsed muon
facility, Rutherford Appleton Laboratory, UK, using the MUSR
spectrometer. Samples with $x$ = 0.0, 0.05, and 0.1 were measured in a
dilution refrigerator at temperatures down to 50~mK and the other in a
standard helium cryostat down to 1.6~K.  The neutron scattering
experiments were performed at the Institut Laue Langevin, Grenoble,
France, on the time-of-flight spectrometer IN6, using an incident energy
of 3.1~meV. Two samples with $x = 0.0$ and $x = 0.05$ were measured in
the dilution fridge down to milliKelvin temperatures.

In a magnetically ordered phase, muons precess in the static internal
field of the sample, which, in zero-field $\mu$SR, leads to well-defined
oscillations of the muon spin depolarization with a frequency that is
proportional to the magnetic order parameter.  In the present
experiment, we modeled the muon depolarization by the sum of two
components, one magnetic and the other nuclear.
\begin{eqnarray}
G(t) & = & A_{m}\left[\frac{2}{3}\cos(2\pi\nu t + \phi)\exp(-\lambda_T t)+
\frac{1}{3}\exp(-\lambda_L t) \right]\nonumber\\ 
& & + A_{n}G_{KT}(t)\label{Eq:1}
\end{eqnarray}
where $A_{m}$ and $A_{n}$ are the magnetic and nuclear amplitudes, $\nu$
is the precession frequency and $\phi$ is its phase, $\lambda_T$ and
$\lambda_L$ are the transverse and longitudinal damping coefficients,
and $G_{KT}(t)$ is the Kubo-Toyabe function that accounts for
depolarization by $^{27}$Al nuclei~\cite{Amato:1997}.  The need for two 
components could mean either that there are two muon sites in these 
samples, one dominated by nuclear and the other by magnetic relaxation, or 
that the samples are magnetically inhomogeneous.  At the lowest 
temperatures, the magnetic oscillations account for about 70\% of the 
depolarization in all the samples we measured, in agreement with earlier 
$\mu$SR results~\cite{Barth:1989} and close to the volume fraction of
magnetically correlated regions derived from NMR
data~\cite{Gavilano:1995}.

\begin{figure}[tbp]
 \vspace{-0.5in}
 \includegraphics[width=3.2in]{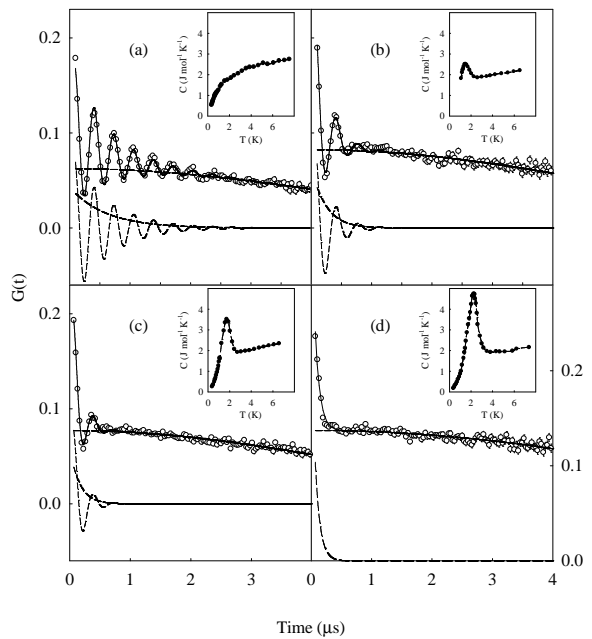}%
 \vspace{-0.5in}
 \caption{Zero field muon spin depolarization in Ce$_{1-x}$La$_x$Al$_3$
 with (a) $x = 0$ at 50~mK, (b) $x = 0.05$ at 100~mK, (c) $x = 0.1$ at
 50~mK, and (d) $x = 0.2$ at 1.6K (note the different scale).  The solid
 lines are fits to Eq. (\ref{Eq:1}), whose three components (transverse
 and longitudinal magnetic, and nuclear) are shown as dashed lines. The
 insets show the specific heat data of Ref.~\cite{Andraka:1995}.
 \label{MuonFig}}
\end{figure}

In our earlier $\mu$SR measurements on samples with $x \geq 0.2$, we did
not observe any oscillations, but Fig.~\ref{MuonFig} shows that they are
indeed present at lower values of $x$.  They are particularly
well-defined in pure CeAl$_3$, but they become increasingly damped with
increasing $x$.  Although our data are in good agreement with earlier
results~\cite{Barth:1989,Amato:1997}, the improved statistical quality
of the present measurements shows that the oscillations follow the
damped sine wave of Eq.~(\ref{Eq:1}), rather than the Bessel function
proposed by Amato~\cite{Amato:1997}.  This indicates that the ordering
does not involve a modulation of the magnitude of the magnetic 
moment~\footnote{Amato's model was designed to explain the origin of the 
large value of $\phi$ seen in all $\mu$SR experiments on CeAl$_3$, 
including our own. This remains an unresolved problem.}.

The longitudinal and transverse damping rates are approximately equal at
temperatures well below T$^*$.  This rate increases dramatically with
$x$, from 1.6~$\mu s^{-1}$ at $x = 0.0$ to 12.5~$\mu s^{-1}$ at $x = 0.2$
so that muon oscillations are no longer observable for $x > 0.1$ and the
magnetic depolarization can be modeled by a simple exponential decay. 
This behavior shows that lanthanum substitution 
suppresses the well-defined magnetic order seen in pure CeAl$_3$,
whereas the specific heat anomalies, shown in the insets to
Fig.~\ref{MuonFig}, become much more pronounced with lanthanum
substitution.  These contrasting trends are the first indication that
the specific heat peaks are not associated with conventional magnetic
phase transitions.

\begin{figure}[tbp]
 \vspace{-0.6in}
 \includegraphics[width=3.2in]{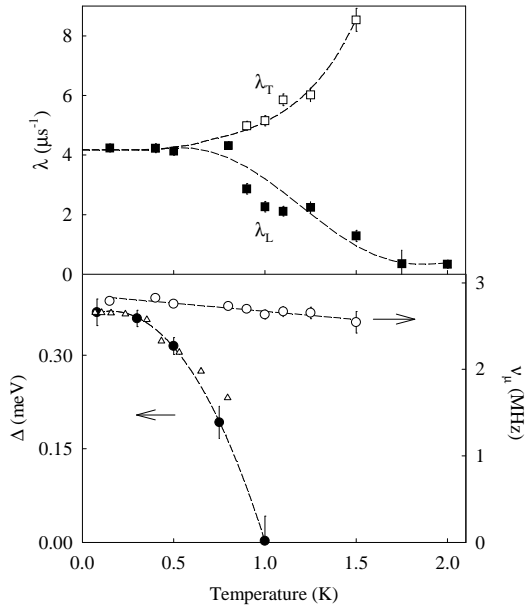}%
 \vspace{-0.5in}
 \caption{Temperature dependence of the muon spin depolarization
 analyzed using Eq. (\ref{Eq:1}) and the excitation energy measured with
 inelastic neutron scattering.  The open and filled squares denote the
 transverse and longitudinal damping coefficients, $\lambda_T$ and
 $\lambda_L$ (below 1~K, $\lambda_T = \lambda_L$), the open circles
 denote the precession frequency, $\nu$, and the filled circles denote
 the excitation energy of the magnetic response, $\Delta$.  The
 triangles are the theoretical values of $\Delta$ derived from an inelastic Lorenzian
 fit to the NRG calculations of S($\omega$).  The dashed lines are guides to the
 eye.
 \label{TFig}}
\end{figure}

This conclusion is strengthened by the detailed temperature dependence
of the muon spin relaxation, which is illustrated by the behavior of $x
= 0.05$ in Fig.~\ref{TFig}.  The most striking result is the nearly
temperature-independent oscillation frequency up to temperatures above
T$^*$.  Therefore, the magnetic order parameter does not change
significantly over the entire temperature range of the specific heat
anomaly, unlike a conventional second-order magnetic phase transition,
in which the order parameter should fall to zero.  Instead, $\lambda_T$
begins to diverge at T$^*$, \textit{i.e.} there is a broadening of
the distribution of internal fields, while $\lambda_L$ starts to fall as
the quasistatic correlations become dynamic. This is consistent with
earlier muon data from pure CeAl$_3$ \cite{Barth:1989,Amato:1997}.

\begin{figure}[tbp]
 \vspace{-0.3in}
 \includegraphics[width=3.2in]{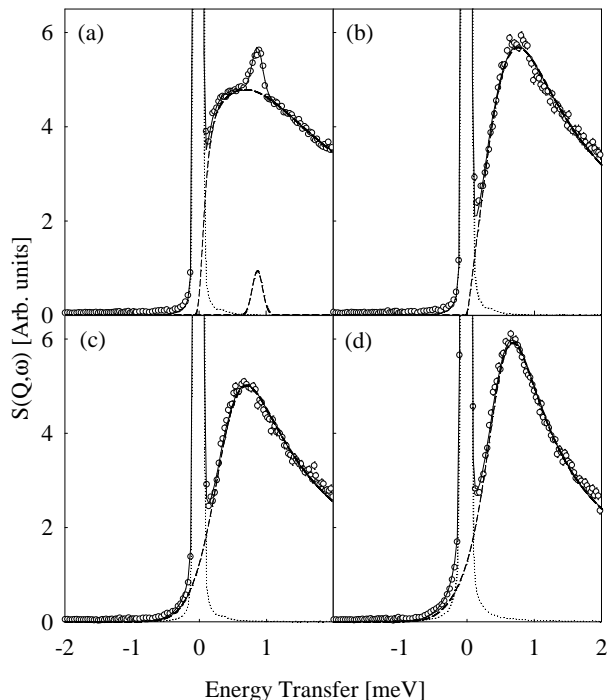}%
 \vspace{-0.5in}
 \caption{Inelastic neutron scattering from Ce$_{1-x}$La$_x$Al$_3$ with
 (a) $x = 0.0$ at T~=~80~mK, (b) $x = 0.05$ at T~=~80~mK, (c) $x = 0.1$
 at T~=~1.6~K, and (d) $x = 0.2$ at T~=~1.6~K.  The solid line is the
 sum of the magnetic response (dashed lines) and the elastic background
(dotted line).
 \label{NeutronFig}}
\end{figure}

We will now compare this trend with the inelastic excitation seen by
neutrons (Fig.~\ref{NeutronFig}).  In pure CeAl$_3$, the response is
almost entirely quasielastic.  There is a small inelastic peak at
0.8~meV that represents only 2\% of the total spectral weight.  It is
well known that there is a strong sample dependence to the properties of
CeAl$_3$, which probably results from an extreme sensitivity to internal
strains~\cite{Brodale:1986}. We believe that the inelastic peak may
arise from small strained regions of the sample and do not discuss it
further.  Although the remaining response is quasielastic, it has a
non-Lorenzian form. In Fig.~\ref{NeutronFig}, we have fitted the data
to the sum of two quasielastic Lorenzian lineshapes (see discussion
later). In all the other samples, the magnetic response is well
described by an Lorenzian lineshape centered at $\pm\Delta$ 
below T$^*$ and a quasielastic Lorenzian lineshape above T$^*$.

Although the magnetic response of pure CeAl$_3$ is quasielastic, it
becomes inelastic with increasing $x$, and the excitation is
progressively better defined at T $\ll$ T$^*$.  At $x=0.05$ (T$^*$ =
0.8~K), $\Delta$ is 0.37(3)~meV and half-width 0.66(2)~meV at 80~mK,
whereas, at $x = 0.2$ (T$^*$ = 2.2~K), $\Delta$ increases to 0.47(1)
meV and and the half-width is 0.42(1)~meV measured at 1.6~K.  The
development of this inelastic response with $x$ correlates well with the
growth of the specific heat peak, in contrast to the behavior of the
muon oscillations. 

Figure \ref{TFig} shows the temperature dependence of
the energy, $\Delta$, of this inelastic excitation.  The fits show that
the energy falls to zero at T$^*$, as if the gap were proportional to an
order parameter. Above T$^*$, the spin dynamics are overdamped.  It is
important to note that the entire magnetic response becomes inelastic
below T$^*$.  There is no evidence of a two-component response as would
be seen if the spin dynamics were spatially inhomogeneous as suggested
by $\mu$SR and NMR~\cite{Barth:1989,Gavilano:1995}. Therefore, although
there is evidence that the static magnetic correlations develop
inhomogeneously, the dynamical transition involves all the cerium ions.

As discussed in Ref.~\cite{Goremychkin:2000}, the excitation cannot
arise from a conventional molecular field splitting of the ground state
doublet, because such a transition is forbidden by dipole selection
rules for a $\lvert\pm\frac{3}{2}\rangle$ Kramers doublet~\cite{Osborn:2000}.
It can only arise if the doublet is split by off-diagonal matrix
elements producing two non-magnetic singlets, as proposed by Rainford
\textit{et al}~\cite{Rainford:1996} in a heuristic analysis of another
anisotropic heavy fermion system, CeRu$_2$Si$_{2-x}$Ge$_x$.  The fact
that $\Delta$ does not track the magnetic moment, which is nearly
temperature independent over the entire temperature range, provides
further confirmation that the excitation is not coupling directly to a
molecular field.

To summarize the experimental conclusions, it is clear that the static
magnetic correlations are not responsible for the thermodynamic
anomalies, but that these anomalies are nevertheless directly correlated
with the opening of a spin gap.  Just such behavior is predicted by the
Anistropic Kondo Model.  The AKM is formally equivalent to a Dissipative
Two-State System or spin boson
model~\cite{Costi:1998,Costi:1999,Chakravarty:1995,Leggett:1987}, in which a
tunneling transition between two singlets is broadened by coupling to
Ohmic dissipation.  When mapped onto the AKM, the transverse Kondo
exchange produces the energy splitting while the axial Kondo exchange
produces the dissipation.  When the anisotropy is sufficiently strong,
defined by the dimensionless parameter $\alpha$ being less than 1/3,
there is a specific heat peak at a characteristic temperature at which
the dynamics cross over from quasielastic to inelastic.  We estimated in
Ref. \cite{Goremychkin:2000} that Ce$_{0.8}$La$_{0.2}$Al$_3$ falls in
this regime with $\alpha\approx 0.1$.  If we apply the same scaling
arguments for the $x$ = 0.05 sample, $\alpha\approx 0.2$, but becomes
0.3, \textit{i.e.}, close to the critical value, at $x$ = 0.0.

Figure~\ref{TFig} also shows the results of a numerical renormalization group (NRG)
calculation of S($\omega$)~\cite{Costi:1996}.  By fitting an inelastic
Lorenzian lineshape to the theoretical calculations as a function of
temperature, we obtain good agreement with the experimental data up to
more than T$^*$/2.  At higher temperatures, as the dynamics become
progressively overdamped, the theoretical lineshape becomes
non-Lorenzian, reminiscent of the unusual lineshape of the
quasielastic response in pure CeAl$_3$. Nevertheless at $x=0.05$, the
agreement is quantitatively good up to 0.7~K and qualitatively describes
the observed transition from inelastic to quasielastic dynamics.

In conclusion, we have observed clear evidence for static magnetic
correlations in $\mu$SR from compounds of Ce$_{1-x}$La$_x$Al$_3$ with $x
\leq 0.1$. However, they become progressively more damped with
increasing $x$ whereas peaks in the specific heat become more
pronounced.  Inelastic neutron scattering shows that these peaks are
associated with the opening of a gap in the single-ion spin dynamics
consistent with the predictions of the AKM, confirming our earlier
conjecture that it is a crossover in the local dynamics rather than
cooperative magnetic ordering that is responsible for the thermodynamic
behavior.

We believe that these conclusions for Ce$_{1-x}$La$_x$Al$_3$ can provide
some insight into the ``hidden order" in URu$_2$Si$_2$
\cite{Shah:2000}.  Both systems have strong Ising-like anisotropy, with
substantial thermodynamic anomalies that are associated with the opening
of a gap in the spin dynamics~\cite{Broholm:1991,Buyers:1994}. There is evidence
that the static magnetic correlations develop inhomogeneously in both
systems~\cite{Matsuda:2001,Gavilano:1995}. We have applied the same AKM
scaling relations as in Ref.~\cite{Goremychkin:2000}, using the
following experimental values for URu$_2$Si$_2$: the electronic specific
heat coefficient and the magnetic susceptibility at T $\rightarrow 0$,
$\gamma$ = 50~mJmol$^{-1}$K$^2$ and
$\chi=2.5\times10^{-3}$~emu~mol$^{-1}$,
respectively~\cite{Palstra:1985}, the excitation energy determined from
inelastic neutron scattering, $\Delta = 5.5$~meV~\cite{Walter:1986}, and
the temperature of the maximum in the specific heat, T$^*$~=~17.5~K.  We
derive $\alpha \approx 0.1$ self-consistently, \textit{i.e.}, it is in
the regime where the AKM predicts a dynamical transition and large
concomitant thermodynamic anomaly.

 It remains to be seen if a lattice version
of the AKM can model the dispersion of the magnetic excitations observed by 
inelastic neutron scattering~\cite{Broholm:1991}. 
Nevertheless, the single-ion model accounts for the main features of 
 URu$_2$Si$_2$ listed above, suggesting that the AKM may form the basis of 
a resolution of the ``hidden order" problem.

\begin{acknowledgments}
We thank G. Zarand and D. E. MacLaughlin for helpful discussions.  This
work was performed under the support of the US Department of Energy, 
Office of Science, under contract no. W-31-109-ENG-38.
\end{acknowledgments}


\begin{thebibliography}{23}
\expandafter\ifx\csname natexlab\endcsname\relax\def\natexlab#1{#1}\fi
\expandafter\ifx\csname bibnamefont\endcsname\relax
  \def\bibnamefont#1{#1}\fi
\expandafter\ifx\csname bibfnamefont\endcsname\relax
  \def\bibfnamefont#1{#1}\fi
\expandafter\ifx\csname citenamefont\endcsname\relax
  \def\citenamefont#1{#1}\fi
\expandafter\ifx\csname url\endcsname\relax
  \def\url#1{\texttt{#1}}\fi
\expandafter\ifx\csname urlprefix\endcsname\relax\def\urlprefix{URL }\fi
\providecommand{\bibinfo}[2]{#2}
\providecommand{\eprint}[2][]{\url{#2}}

\bibitem[{\citenamefont{Andres et~al.}(1975)\citenamefont{Andres, Graebner, and
  Ott}}]{Andres:1975}
\bibinfo{author}{\bibfnamefont{K.}~\bibnamefont{Andres}},
  \bibinfo{author}{\bibfnamefont{J.~E.} \bibnamefont{Graebner}},
  \bibnamefont{and} \bibinfo{author}{\bibfnamefont{H.~R.} \bibnamefont{Ott}},
  \bibinfo{journal}{Phys.\ Rev.\ Lett.} \textbf{\bibinfo{volume}{35}},
  \bibinfo{pages}{1779} (\bibinfo{year}{1975}).

\bibitem[{\citenamefont{Bredl et~al.}(1984)\citenamefont{Bredl, Horn, Steglich,
  L{\"u}thi, and Martin}}]{Bredl:1984}
\bibinfo{author}{\bibfnamefont{C.~D.} \bibnamefont{Bredl}},
  \bibinfo{author}{\bibfnamefont{S.}~\bibnamefont{Horn}},
  \bibinfo{author}{\bibfnamefont{F.}~\bibnamefont{Steglich}},
  \bibinfo{author}{\bibfnamefont{B.}~\bibnamefont{L{\"u}thi}},
  \bibnamefont{and} \bibinfo{author}{\bibfnamefont{R.~M.}
  \bibnamefont{Martin}}, \bibinfo{journal}{Phys.\ Rev.\ Lett.}
  \textbf{\bibinfo{volume}{52}}, \bibinfo{pages}{1982} (\bibinfo{year}{1984}).

\bibitem[{\citenamefont{Barth et~al.}(1989)\citenamefont{Barth, Ott, Gygax,
  Hitti, Lippelt, Schenck, and Baines}}]{Barth:1989}
\bibinfo{author}{\bibfnamefont{S.}~\bibnamefont{Barth}},
  \bibinfo{author}{\bibfnamefont{H.~R.} \bibnamefont{Ott}},
  \bibinfo{author}{\bibfnamefont{F.~N.} \bibnamefont{Gygax}},
  \bibinfo{author}{\bibfnamefont{B.}~\bibnamefont{Hitti}},
  \bibinfo{author}{\bibfnamefont{E.}~\bibnamefont{Lippelt}},
  \bibinfo{author}{\bibfnamefont{A.}~\bibnamefont{Schenck}}, \bibnamefont{and}
  \bibinfo{author}{\bibfnamefont{C.}~\bibnamefont{Baines}},
  \bibinfo{journal}{Phys.\ Rev.\ B} \textbf{\bibinfo{volume}{39}},
  \bibinfo{pages}{11695} (\bibinfo{year}{1989}).

\bibitem[{\citenamefont{Wong and Clark}(1992)}]{Wong:1992}
\bibinfo{author}{\bibfnamefont{W.~H.} \bibnamefont{Wong}} \bibnamefont{and}
  \bibinfo{author}{\bibfnamefont{W.~G.} \bibnamefont{Clark}},
  \bibinfo{journal}{J.\ Magn.\ Magn.\ Mater.} \textbf{\bibinfo{volume}{108}},
  \bibinfo{pages}{175} (\bibinfo{year}{1992}).

\bibitem[{\citenamefont{Gavilano et~al.}(1995)\citenamefont{Gavilano, Hunziker,
  and Ott}}]{Gavilano:1995}
\bibinfo{author}{\bibfnamefont{J.~L.} \bibnamefont{Gavilano}},
  \bibinfo{author}{\bibfnamefont{J.}~\bibnamefont{Hunziker}}, \bibnamefont{and}
  \bibinfo{author}{\bibfnamefont{H.~R.} \bibnamefont{Ott}},
  \bibinfo{journal}{Phys.\ Rev.\ B} \textbf{\bibinfo{volume}{52}},
  \bibinfo{pages}{R13106} (\bibinfo{year}{1995}).

\bibitem[{\citenamefont{Shah et~al.}(2000)\citenamefont{Shah, Chandra, Coleman,
  and Mydosh}}]{Shah:2000}
\bibinfo{author}{\bibfnamefont{N.}~\bibnamefont{Shah}},
  \bibinfo{author}{\bibfnamefont{P.}~\bibnamefont{Chandra}},
  \bibinfo{author}{\bibfnamefont{P.}~\bibnamefont{Coleman}}, \bibnamefont{and}
  \bibinfo{author}{\bibfnamefont{J.~A.} \bibnamefont{Mydosh}},
  \bibinfo{journal}{Phys.\ Rev.\ B} \textbf{\bibinfo{volume}{61}},
  \bibinfo{pages}{564} (\bibinfo{year}{2000}).

\bibitem[{\citenamefont{Andraka et~al.}(1995)\citenamefont{Andraka, Jee, and
  Stewart}}]{Andraka:1995}
\bibinfo{author}{\bibfnamefont{B.}~\bibnamefont{Andraka}},
  \bibinfo{author}{\bibfnamefont{C.~S.} \bibnamefont{Jee}}, \bibnamefont{and}
  \bibinfo{author}{\bibfnamefont{G.~R.} \bibnamefont{Stewart}},
  \bibinfo{journal}{Phys.\ Rev.\ B} \textbf{\bibinfo{volume}{52}},
  \bibinfo{pages}{9462} (\bibinfo{year}{1995}).

\bibitem[{\citenamefont{Goremychkin et~al.}(2000)\citenamefont{Goremychkin,
  Osborn, Rainford, and Murani}}]{Goremychkin:2000}
\bibinfo{author}{\bibfnamefont{E.~A.} \bibnamefont{Goremychkin}},
  \bibinfo{author}{\bibfnamefont{R.}~\bibnamefont{Osborn}},
  \bibinfo{author}{\bibfnamefont{B.~D.} \bibnamefont{Rainford}},
  \bibnamefont{and} \bibinfo{author}{\bibfnamefont{A.~P.}
  \bibnamefont{Murani}}, \bibinfo{journal}{Phys.\ Rev.\ Lett.}
  \textbf{\bibinfo{volume}{84}}, \bibinfo{pages}{2211} (\bibinfo{year}{2000}).

\bibitem[{\citenamefont{Costi}(1998)}]{Costi:1998}
\bibinfo{author}{\bibfnamefont{T.~A.} \bibnamefont{Costi}},
  \bibinfo{journal}{Phys.\ Rev.\ Lett.} \textbf{\bibinfo{volume}{80}},
  \bibinfo{pages}{1038} (\bibinfo{year}{1998}).

\bibitem[{\citenamefont{Pietri et~al.}(2001)\citenamefont{Pietri, Ingersent,
  and Andraka}}]{Pietri:2001}
\bibinfo{author}{\bibfnamefont{R.}~\bibnamefont{Pietri}},
  \bibinfo{author}{\bibfnamefont{K.}~\bibnamefont{Ingersent}},
  \bibnamefont{and} \bibinfo{author}{\bibfnamefont{B.}~\bibnamefont{Andraka}},
  \bibinfo{journal}{Phys.\ Rev.\ Lett.} \textbf{\bibinfo{volume}{86}},
  \bibinfo{pages}{1090} (\bibinfo{year}{2001}).

\bibitem[{\citenamefont{Matsuda et~al.}(2001)\citenamefont{Matsuda, Kohori,
  Kohara, Kuwahara, and Amitsuka}}]{Matsuda:2001}
\bibinfo{author}{\bibfnamefont{K.}~\bibnamefont{Matsuda}},
  \bibinfo{author}{\bibfnamefont{Y.}~\bibnamefont{Kohori}},
  \bibinfo{author}{\bibfnamefont{T.}~\bibnamefont{Kohara}},
  \bibinfo{author}{\bibfnamefont{K.}~\bibnamefont{Kuwahara}}, \bibnamefont{and}
  \bibinfo{author}{\bibfnamefont{H.}~\bibnamefont{Amitsuka}},
  \bibinfo{journal}{Phys.\ Rev.\ Lett.} \textbf{\bibinfo{volume}{8708}},
  \bibinfo{pages}{7203} (\bibinfo{year}{2001}).

\bibitem[{\citenamefont{Amato}(1997)}]{Amato:1997}
\bibinfo{author}{\bibfnamefont{A.}~\bibnamefont{Amato}},
  \bibinfo{journal}{Rev.\ Mod.\ Phys.} \textbf{\bibinfo{volume}{69}},
  \bibinfo{pages}{1119} (\bibinfo{year}{1997}).

\bibitem[{\citenamefont{Brodale et~al.}(1986)\citenamefont{Brodale, Fisher,
  Phillips, and Flouquet}}]{Brodale:1986}
\bibinfo{author}{\bibfnamefont{G.~E.} \bibnamefont{Brodale}},
  \bibinfo{author}{\bibfnamefont{R.~A.} \bibnamefont{Fisher}},
  \bibinfo{author}{\bibfnamefont{N.~E.} \bibnamefont{Phillips}},
  \bibnamefont{and} \bibinfo{author}{\bibfnamefont{J.}~\bibnamefont{Flouquet}},
  \bibinfo{journal}{Phys.\ Rev.\ Lett.} \textbf{\bibinfo{volume}{56}},
  \bibinfo{pages}{390} (\bibinfo{year}{1986}).

\bibitem[{\citenamefont{Osborn et~al.}(2000)\citenamefont{Osborn, Goremychkin,
  Rainford, Sashin, and Murani}}]{Osborn:2000}
\bibinfo{author}{\bibfnamefont{R.}~\bibnamefont{Osborn}},
  \bibinfo{author}{\bibfnamefont{E.~A.} \bibnamefont{Goremychkin}},
  \bibinfo{author}{\bibfnamefont{B.~D.} \bibnamefont{Rainford}},
  \bibinfo{author}{\bibfnamefont{I.~L.} \bibnamefont{Sashin}},
  \bibnamefont{and} \bibinfo{author}{\bibfnamefont{A.~P.}
  \bibnamefont{Murani}}, \bibinfo{journal}{J.\ Appl.\ Phys.}
  \textbf{\bibinfo{volume}{87}}, \bibinfo{pages}{5131} (\bibinfo{year}{2000}).

\bibitem[{\citenamefont{Rainford et~al.}(1996)\citenamefont{Rainford, Neville,
  Adroja, Dakin, and Murani}}]{Rainford:1996}
\bibinfo{author}{\bibfnamefont{B.~D.} \bibnamefont{Rainford}},
  \bibinfo{author}{\bibfnamefont{A.~J.} \bibnamefont{Neville}},
  \bibinfo{author}{\bibfnamefont{D.~T.} \bibnamefont{Adroja}},
  \bibinfo{author}{\bibfnamefont{S.~J.} \bibnamefont{Dakin}}, \bibnamefont{and}
  \bibinfo{author}{\bibfnamefont{A.~P.} \bibnamefont{Murani}},
  \bibinfo{journal}{Physica B} \textbf{\bibinfo{volume}{224}},
  \bibinfo{pages}{163} (\bibinfo{year}{1996}).

\bibitem[{\citenamefont{Costi and Zar{\'a}nd}(1999)}]{Costi:1999}
\bibinfo{author}{\bibfnamefont{T.~A.} \bibnamefont{Costi}} \bibnamefont{and}
  \bibinfo{author}{\bibfnamefont{G.}~\bibnamefont{Zar{\'a}nd}},
  \bibinfo{journal}{Phys.\ Rev.\ B} \textbf{\bibinfo{volume}{59}},
  \bibinfo{pages}{12398} (\bibinfo{year}{1999}).

\bibitem[{\citenamefont{Chakravarty and Rudnick}(1995)}]{Chakravarty:1995}
\bibinfo{author}{\bibfnamefont{S.}~\bibnamefont{Chakravarty}} \bibnamefont{and}
  \bibinfo{author}{\bibfnamefont{J.}~\bibnamefont{Rudnick}},
  \bibinfo{journal}{Phys.\ Rev.\ Lett.} \textbf{\bibinfo{volume}{75}},
  \bibinfo{pages}{591} (\bibinfo{year}{1995}).

\bibitem[{\citenamefont{Leggett et~al.}(1987)\citenamefont{Leggett,
  Chakravarty, Dorsey, Fisher, Garg, and Zwerger}}]{Leggett:1987}
\bibinfo{author}{\bibfnamefont{A.~J.} \bibnamefont{Leggett}},
  \bibinfo{author}{\bibfnamefont{S.}~\bibnamefont{Chakravarty}},
  \bibinfo{author}{\bibfnamefont{A.~T.} \bibnamefont{Dorsey}},
  \bibinfo{author}{\bibfnamefont{M.~P.~A.} \bibnamefont{Fisher}},
  \bibinfo{author}{\bibfnamefont{A.}~\bibnamefont{Garg}}, \bibnamefont{and}
  \bibinfo{author}{\bibfnamefont{W.}~\bibnamefont{Zwerger}},
  \bibinfo{journal}{Rev.\ Mod.\ Phys.} \textbf{\bibinfo{volume}{59}},
  \bibinfo{pages}{1} (\bibinfo{year}{1987}).

\bibitem[{\citenamefont{Costi and Kieffer}(1996)}]{Costi:1996}
\bibinfo{author}{\bibfnamefont{T.~A.} \bibnamefont{Costi}} \bibnamefont{and}
  \bibinfo{author}{\bibfnamefont{C.}~\bibnamefont{Kieffer}},
  \bibinfo{journal}{Phys.\ Rev.\ Lett.} \textbf{\bibinfo{volume}{76}},
  \bibinfo{pages}{1683} (\bibinfo{year}{1996}).

\bibitem[{\citenamefont{Broholm et~al.}(1991)\citenamefont{Broholm, Lin,
  Matthews, Mason, Buyers, Collins, Menovsky, Mydosh, and
  Kjems}}]{Broholm:1991}
\bibinfo{author}{\bibfnamefont{C.}~\bibnamefont{Broholm}},
  \bibinfo{author}{\bibfnamefont{H.}~\bibnamefont{Lin}},
  \bibinfo{author}{\bibfnamefont{P.~T.} \bibnamefont{Matthews}},
  \bibinfo{author}{\bibfnamefont{T.~E.} \bibnamefont{Mason}},
  \bibinfo{author}{\bibfnamefont{W.~J.~L.} \bibnamefont{Buyers}},
  \bibinfo{author}{\bibfnamefont{M.~F.} \bibnamefont{Collins}},
  \bibinfo{author}{\bibfnamefont{A.~A.} \bibnamefont{Menovsky}},
  \bibinfo{author}{\bibfnamefont{J.~A.} \bibnamefont{Mydosh}},
  \bibnamefont{and} \bibinfo{author}{\bibfnamefont{J.~K.} \bibnamefont{Kjems}},
  \bibinfo{journal}{Phys.\ Rev.\ B} \textbf{\bibinfo{volume}{43}},
  \bibinfo{pages}{12809} (\bibinfo{year}{1991}).

\bibitem[{\citenamefont{Buyers et~al.}(1994)\citenamefont{Buyers, Tun,
  Petersen, Mason, Lussier, Gaulin, and Menovsky}}]{Buyers:1994}
\bibinfo{author}{\bibfnamefont{W.~J.~L.} \bibnamefont{Buyers}},
  \bibinfo{author}{\bibfnamefont{Z.}~\bibnamefont{Tun}},
  \bibinfo{author}{\bibfnamefont{T.}~\bibnamefont{Petersen}},
  \bibinfo{author}{\bibfnamefont{T.~E.} \bibnamefont{Mason}},
  \bibinfo{author}{\bibfnamefont{J.-G.} \bibnamefont{Lussier}},
  \bibinfo{author}{\bibfnamefont{B.~D.} \bibnamefont{Gaulin}},
  \bibnamefont{and} \bibinfo{author}{\bibfnamefont{A.~A.}
  \bibnamefont{Menovsky}}, \bibinfo{journal}{Physica\ B}
  \textbf{\bibinfo{volume}{199 \& 200}}, \bibinfo{pages}{95}
  (\bibinfo{year}{1994}).

\bibitem[{\citenamefont{Palstra et~al.}(1985)\citenamefont{Palstra, Menovsky,
  van~den Berg, Dirkmaat, Kes, Nieuwenhuys, and Mydosh}}]{Palstra:1985}
\bibinfo{author}{\bibfnamefont{T.~T.~M.} \bibnamefont{Palstra}},
  \bibinfo{author}{\bibfnamefont{A.~A.} \bibnamefont{Menovsky}},
  \bibinfo{author}{\bibfnamefont{J.}~\bibnamefont{van~den Berg}},
  \bibinfo{author}{\bibfnamefont{A.~J.} \bibnamefont{Dirkmaat}},
  \bibinfo{author}{\bibfnamefont{P.~H.} \bibnamefont{Kes}},
  \bibinfo{author}{\bibfnamefont{G.~J.} \bibnamefont{Nieuwenhuys}},
  \bibnamefont{and} \bibinfo{author}{\bibfnamefont{J.~A.}
  \bibnamefont{Mydosh}}, \bibinfo{journal}{Phys.\ Rev.\ Lett.}
  \textbf{\bibinfo{volume}{55}}, \bibinfo{pages}{2727} (\bibinfo{year}{1985}).

\bibitem[{\citenamefont{Walter et~al.}(1986)\citenamefont{Walter, Loong,
  Loewenhaupt, and Schlabitz}}]{Walter:1986}
\bibinfo{author}{\bibfnamefont{U.}~\bibnamefont{Walter}},
  \bibinfo{author}{\bibfnamefont{C.-K.} \bibnamefont{Loong}},
  \bibinfo{author}{\bibfnamefont{M.}~\bibnamefont{Loewenhaupt}},
  \bibnamefont{and}
  \bibinfo{author}{\bibfnamefont{W.}~\bibnamefont{Schlabitz}},
  \bibinfo{journal}{Phys.\ Rev.\ B} \textbf{\bibinfo{volume}{33}},
  \bibinfo{pages}{7875} (\bibinfo{year}{1986}).

\end{thebibliography}

\end{document}